\documentclass{article}

\usepackage[preprint, nonatbib]{neurips_2024}

\usepackage{cite}
\usepackage[english]{babel}
\usepackage{amsmath,graphicx}
\usepackage{amssymb}
\usepackage{amsmath}
\usepackage{amsthm}
\usepackage{pgfplots}
\usepackage{float}
\usepackage{colortbl}
\usepackage[pagebackref=true,breaklinks=true,colorlinks,bookmarks=false]{hyperref}

\usepackage[utf8]{inputenc} 
\usepackage[T1]{fontenc}    
\usepackage{hyperref}       
\usepackage{url}            
\usepackage{booktabs}       
\usepackage{amsfonts}       
\usepackage{nicefrac}       
\usepackage{microtype}      
\usepackage{xcolor}         

\usepackage{tabularx}
\usepackage{multirow}
\usepackage{multicol}
\usepackage{array,booktabs,caption,ragged2e}

\def\*#1{\mathbf{#1}}

\newcolumntype{C}{>{\centering\arraybackslash}X}%
\newcolumntype{L}{>{\raggedleft\arraybackslash}X}%
\newcolumntype{R}{>{\raggedright\arraybackslash}X}%

\newcolumntype{D}[1]{>{\hsize=#1\hsize\centering\arraybackslash}X}%
\newcolumntype{T}[1]{>{\hsize=#1\hsize\raggedright\arraybackslash}X}%
\newcolumntype{M}[1]{>{\hsize=#1\hsize\raggedleft\arraybackslash}X}%

\newcolumntype{Y}[2]{%
    >{\adjustbox{angle=#1,lap=\width-(#2)}\bgroup}%
    l%
    <{\egroup}%
}

\def\jetOnera{JET }

\def\ie{{\textit i.e.~}}

\definecolor{orange}{rgb}{1.0,0.5,0}

\begin{document}

\title{Evaluating the Posterior Sampling Ability of Plug\&Play Diffusion Methods in Sparse-View CT}

\author{
  \hspace{-0.7cm}Liam Moroy$~^{1,2}$ \quad Guillaume Bourmaud$~^{2}$ \quad Frédéric Champagnat$~^{1}$ \quad  Jean-François Giovannelli$~^{2}$ \vspace{0.15cm}\\ 
  $^1$ DTIS/ONERA, Univ. Paris Saclay, France,\\$^2$ IMS (Univ. Bordeaux~--~CNRS~--~BINP), France
}

\maketitle

\begin{abstract}
  Plug\&Play (PnP) diffusion models are state-of-the-art methods in computed tomography (CT) reconstruction. Such methods usually consider applications where the sinogram contains a sufficient amount of information for the posterior distribution to be concentrated around a single mode, and consequently are evaluated using image-to-image metrics such as PSNR/SSIM. 
Instead, we are interested in reconstructing compressible flow
images from sinograms having a small number of projections, which results in a posterior distribution no longer concentrated or even multimodal.
Thus, in this paper, we aim at evaluating the approximate posterior of PnP diffusion models and introduce two posterior evaluation properties.
We quantitatively evaluate three PnP diffusion methods on three different datasets for several numbers of projections. We surprisingly find that, for each method, the approximate posterior deviates from the true posterior when the number of projections decreases.

\end{abstract}

%
%
\section{Introduction}
\label{sec:intro}

\noindent Diffusion models learn the prior of an underlying data distribution, allowing to generate new samples~\cite{sohl2015deep, ho2020denoising, dhariwal2021diffusion, song2021score, kingma2021variational, song2019generative}. Plug\&Play (PnP) diffusion models~\cite{song2021score, kawar2021snips, kawar2022denoising, song2022solving, chung2022improving, chung2023diffusion, song2023pseudoinverse, choi2021ilvr} employ such prior-encoding models to solve inverse problems.
 In this paper, we focus on the Sparse-View Computed Tomography (SVCT)~\cite{kak2001principles} measurement model :
\begin{equation}\label{eq:observation_model}
    \mathbf{y}_p = \mathbf{H}_p\mathbf{x} + \boldsymbol{\epsilon}_{p},
\end{equation}
where $\mathbf{y}_p \in \mathbb{R}^{m_p}$ is the measured sinogram with $p$ projections, $\mathbf{H}_p$ is a discretized Radon transform matrix of size $m_p \times n$ corresponding to a parallel beam setting, $\mathbf{x} \in \mathbb{R}^n$ is the image and $\boldsymbol{\epsilon}_{p} \sim \mathcal{N}(\mathbf{0}, \sigma_y^2\mathbf{I}_{m_p})$ is the measurement noise. 

At test time, only $\mathbf{y}_p$, $\mathbf{H}_p$, $\sigma_y^2$ and the prior-encoding diffusion model are known. In practice, $\mathbf{H}_p$ and $\sigma_y^2$ depend on the tomograph settings. For this reason, we focus on PnP diffusion models that avoid any extra learning stage specific to the measurement model and/or its parameters, as opposed to conditional diffusion models~\cite{saharia2022image, xie2022measurement, ho2021classifier}.

\begin{figure}[t]
  \centering
  \includegraphics[width=10cm]{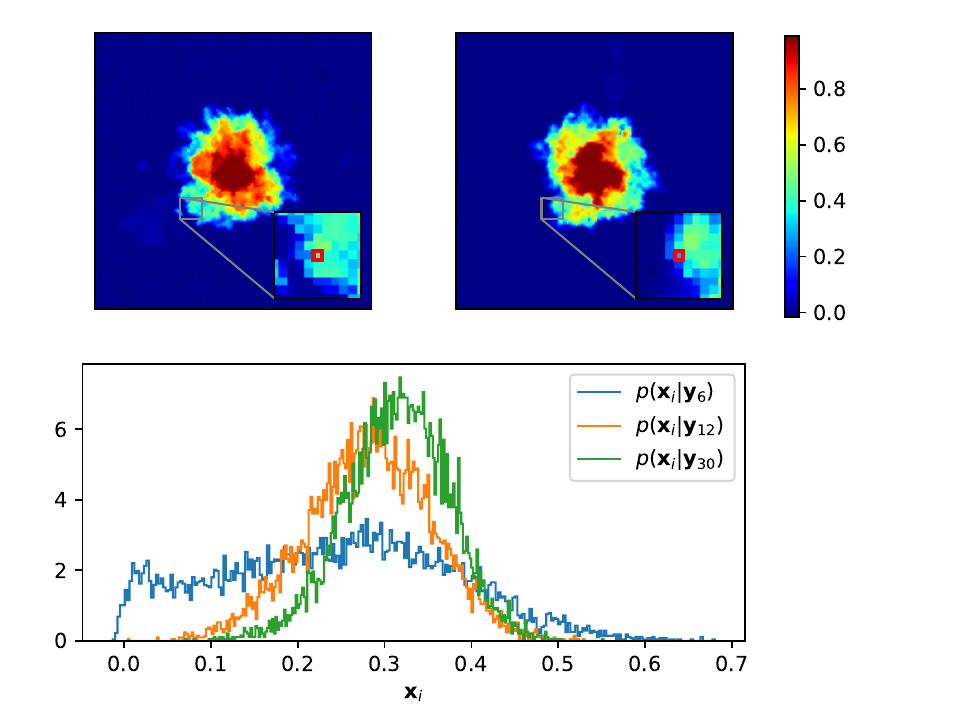}
  \captionof{figure}{
Top row: two compressible flow posterior samples from a PnP diffusion model~\cite{chung2023diffusion} using the same measured sinogram $\mathbf{y}_6$, \ie $p\text{\,=\,}6$. The two samples look very different because the posterior is not peaked when the sinogram contains few projections. 
Bottom: histograms of 10000 posterior samples~\cite{chung2023diffusion} of the red boxed pixel location $\mathbf{x}_i$, for $p\text{\,=\,}6, 12 \text{ and } 30$ projections. 
When $p\text{\,=\,}6$ the posterior is not peaked anymore, thus a point estimation of $\mathbf{x}$ is not sufficient and being able to properly sample from the true posterior is fundamental, which we evaluate in this paper. 
\vspace{-0.5cm}}\label{fig:JetSamples}
\end{figure}

The majority of PnP diffusion models for SVCT~\cite{wu2023data, du2024dper, guan2023generative, chung2022improving, song2022solving} consider applications where the measured sinogram contains a sufficient amount of information for the posterior distribution $p(\mathbf{x} \vert \mathbf{y}_p)$ to be \textit{peaked}, \ie concentrated around a single mode (see Fig.~\ref{fig:JetSamples}). In this case, a point estimation of $\mathbf{x}$ is sufficient to evaluate such posterior and consequently image-to-image metrics such as PSNR/SSIM~\cite{wang2004image} may be employed to evaluate the performances of a given approach.
Instead, this paper is motivated by the reconstruction of compressible jet flows images from a small number of projections~\cite{leon22dhi}, whether for limited optical access or complexity of the setup, as it is the case of digital holographic interferometry~\cite{leon22dhi}, where $p=6$.
In this context, the posterior distribution $p(\mathbf{x} \vert \mathbf{y}_p)$ is no longer peaked and may even be multimodal (see Fig.~\ref{fig:JetSamples}). In this case, a point estimation of $\mathbf{x}$ is no longer sufficient to evaluate such posterior, therefore sampling the true posterior distribution $p(\mathbf{x} \vert \mathbf{y}_p)$ is fundamental.
PnP diffusion methods are approximate posterior samplers. However, to the best of our knowledge, their ability to sample the true posterior, in SVCT, has not been evaluated as of yet.

The contributions of the paper are as follows:\\
(i) We derive two properties that a ``good" approximate posterior should satisfy, one in image space and the other in measurement space.\\
(ii) We quantitatively evaluate three PnP diffusion methods on three different datasets for several numbers of projections, ranging from $p=1$ to $p=180$ projections.\\
(iii) We surprisingly find that, for each method, the approximate posterior deviates from the true posterior when the number of projections decreases.

\section{Related work}
\label{sec:related_work}
\noindent This work focuses on evaluating the posterior sampling ability of different PnP diffusion methods on SVCT when the posterior is not peaked or even multimodal. 

State-of-the-art (SOTA) SVCT methods~\cite{chung2023diffusion, chung2022improving, wu2023data, du2024dper, sun2021coil, wu2023self, song2022solving, jin2017deep} are essentially developed and evaluated to deal with cases where the posterior is peaked, which often corresponds to a large number of projections. Let us highlight that the term Ultra Sparse-View CT (USVCT)~\cite{tan2024msdiff, du2024dper, wu2023data, Li_2023_ICCV} has emerged to perform CT reconstruction from highly sparse sinograms. However, these works are still limited to sinograms with at least $10$ projections, a regime where the posterior is, in the considered applications, still peaked. To the best of our knowledge, this paper is the first attempt to evaluate PnP diffusion methods where the measured sinogram does not contain sufficient information for the posterior distribution to be peaked (see Fig.\ref{fig:JetSamples}).

The above methods report their results in terms of PSNR and/or SSIM~\cite{wang2004image}. As these metrics focus on image-to-image comparisons, a true posterior sample may obtain a low score when the posterior is not peaked. Thus, they are not well suited to evaluate the ability of a given method to sample the true posterior. In some other inverse problems, \textit{e.g.} inpainting~\cite{lugmayr2022repaint, chung2023diffusion, song2023pseudoinverse}, super-resolution~\cite{saharia2022image, song2023pseudoinverse}, generative metrics such as \textit{Fréchet Inception Distance} (FID)~\cite{heusel2017gans} and \textit{Kernel Inception Distance} (KID)~\cite{binkowski2018demystifying} are used to evaluate the distribution of reconstructed samples. To the best of our knowledge, such generative metrics have not yet been employed in SVCT and this paper is the first attempt to do so.

\section{Background}
 \noindent \textbf{Diffusion models}~\cite{ho2020denoising, song2021score, dhariwal2021diffusion, song2019generative} aim at sampling from a prior distribution $p_0$, by defining a generative process that gradually transforms a sample from a known distribution $p_T$ into a sample from $p_0$. To do so, the following reverse-time Stochastic Differential Equation (SDE)~\cite{song2021score} is solved:
\begin{equation}\label{eq:reverseTimeSDE}
    \text{d}\mathbf{x} = \left[ \mathbf{f}(\mathbf{x}, t) - g^2(t)\nabla_{\mathbf{x}}\log p_t(\mathbf{x}) \right]\text{d}t + g(t)\text{d}\bar{\mathbf{w}},
\end{equation}
where $\mathbf{f}(\mathbf{x}, t)$ is the drift function, $g(t)$ is the diffusion coefficient, $\bar{\mathbf{w}}$ is the standard Wiener process flowing backward in time, and $p_t$ is a perturbed version of the prior distribution $p_0$: $p_t(\mathbf{x}') = \int p_t(\mathbf{x}' \vert \mathbf{x})p_0(\mathbf{x})\text{d}\mathbf{x}$, with $ p_t(\mathbf{x}' \vert \mathbf{x})$ being the perturbation kernel. Throughout the paper, we consider the variance-exploding (VE) SDE settings\footnote{ VE-SDE settings: $\textbf{f}(\mathbf{x}, t) = \mathbf{0}$, $g^2(t) = \frac{\text{d}}{\text{d}t}\sigma^2(t)$, $p_t(\mathbf{x}' \vert \mathbf{x}) = \mathcal{N}(\mathbf{x}, \sigma^2(t)\mathbf{I})$ and $\sigma(t) = \sigma_{\text{min}}(\sigma_{\text{max}} / \sigma_{\text{min}})^t$}.
In practice, the time-dependent prior score function $\nabla_{\mathbf{x}}\log p_t(\mathbf{x})$ is approximated by a neural network $\mathbf{s}_{\boldsymbol{\theta}}(\mathbf{x}, t)$ trained over a set of $K$ example images $\left\{\mathbf{x}^{(k)}\right\}_{k=1}^{K}$ 
using \textit{Denoising Score-Matching}~\cite{vincent2011connection}. In the rest of the paper, we will refer to $\mathbf{s}_{\boldsymbol{\theta}}$ as the time-dependent prior score network. Thus, to (approximately) sample from $p_0$, the reverse-time SDE Eq.~\eqref{eq:reverseTimeSDE} is simulated with $\mathbf{s}_{\boldsymbol{\theta}}(\mathbf{x}, t)$ instead of $\nabla_{\mathbf{x}}\log p_t(\mathbf{x})$.\\
\noindent \textbf{PnP diffusion models}~\cite{chung2023diffusion, chung2022improving, song2023pseudoinverse} aim at sampling from the posterior distribution $p_0(\mathbf{x} \vert \mathbf{y})$ while avoiding any extra learning stage specific to the measurement model and/or its parameters. 

To do so, the following reverse-time SDE is employed:
\begin{equation}
    \text{d}\mathbf{x} = \left[ \mathbf{f}(\mathbf{x}, t) - g^2(t)\nabla_{\mathbf{x}}\log p_t(\mathbf{x} \vert \mathbf{y}) \right]\text{d}t + g(t)\text{d}\bar{\mathbf{w}},
\end{equation}
and the time-dependent posterior score function is decomposed as follows:
\begin{align}
    \nabla_{\mathbf{x}}\log p_t(\mathbf{x} \vert \mathbf{y}) &= \underbrace{\nabla_{\mathbf{x}}\log p_t(\mathbf{x})}_{\textstyle \approx \,\mathbf{s}_{\boldsymbol{\theta}}(\mathbf{x}_t, t)} + \nabla_{\mathbf{x}}\log p_t( \mathbf{y} \vert \mathbf{x}).\label{eq:Bayes}
\end{align}
 Because of its dependence w.r.t. $t$, $\nabla_{\mathbf{x}}\log p_t( \mathbf{y} \vert \mathbf{x})$ is not tractable. To circumvent this problem,~\cite{chung2022improving, chung2022come} applies projections onto the measurement subset,~\cite{song2022solving, du2024dper, liu2023dolce} solve an optimization problem and~\cite{chung2023diffusion, song2023pseudoinverse, chung2022improving} propose analytical approximations. From this point of view, PnP diffusion methods are``approximately" true
posterior samplers. In this paper, we focus on evaluating how good these approximations are, on SVCT, and consider three SOTA methods MCG~\cite{chung2022improving}, DPS~\cite{chung2023diffusion}, and $\Pi$G~\cite{song2023pseudoinverse}.
Their analytical approximations of $\nabla_{\mathbf{x}}\log p_t( \mathbf{y} \vert \mathbf{x})$ are summarized in Tab.~\ref{tab:grad}.

It is worth noting that, in the inverse problem literature, several methods \cite{song2023pseudoinverse, gungor2024bayesian} are referred to as \emph{problem agnostic} since the inversion relies on a network whose training is independent of the observation model. These \emph{problem agnostic} methods are distinct from so-called \emph{agnostic methods} \cite{gilton2019neumann}, which train a network to directly map the observations $\mathbf{y}_p$ to their ground truth $\mathbf{x}$; in this context, \emph{agnostic} means that the observation model is not explicitly defined but implicitly represented by the pairs $(\mathbf{y}_p, \mathbf{x})$.
\begin{table}[t]
    \begin{tabularx}{1.0\linewidth}{D{0.15}D{0.85}}
Method &  Approx. of $\nabla_{\mathbf{x}_t}\log p_t( \mathbf{y}_p \vert \mathbf{x}_t)$\\[5pt] \toprule
       MCG \cite{chung2022improving} & $\alpha(\mathbf{x}_t, \mathbf{y}_p)\frac{\partial\hat{\mathbf{x}}_0(\mathbf{x}_t)}{\partial\mathbf{x}_t}\mathbf{H}_p^{\dag}(\mathbf{y}_p - \mathbf{H}_p\hat{\mathbf{x}}_0(\mathbf{x}_t))$\\[5pt]\midrule
       DPS \cite{chung2023diffusion} & $\alpha(\mathbf{x}_t, \mathbf{y}_p)\frac{\partial\hat{\mathbf{x}}_0(\mathbf{x}_t)}{\partial\mathbf{x}_t} \mathbf{H}_p^\top(\mathbf{y}_p - \mathbf{H}_p\hat{\mathbf{x}}_0(\mathbf{x}_t))$\\[5pt]\midrule
       $\Pi$G \cite{song2023pseudoinverse} & $\frac{\partial\hat{\mathbf{x}}_0(\mathbf{x}_t)}{\partial\mathbf{x}_t}\mathbf{H}_p^\top(r_t^2\mathbf{H}_p\mathbf{H}_p^\top + \sigma^2_y\mathbf{I})^{-1}(\mathbf{y}_p - \mathbf{H}_p\hat{\mathbf{x}}_0(\mathbf{x}_t))$\\[5pt]\bottomrule
    \end{tabularx}
    \vspace{0.2cm}
    \caption{Approximations of $\nabla_{\mathbf{x}_t}\log p_t( \mathbf{y}_p \vert \mathbf{x}_t)$ for three SOTA PnP diffusion methods on the SVCT measurement model (Eq.~\eqref{eq:observation_model}). In the table, $\alpha_{\{\text{DPS}, \text{MCG}\}}(\mathbf{x}_t, \mathbf{y}_p)$ are handcrafted weights (see Sec.~\ref{sec:technical_details_pnp}), $\hat{\mathbf{x}}_0(\mathbf{x}_t)= \mathbf{x}_t + \sigma_t^2 \mathbf{s}_{\boldsymbol{\theta}}(\mathbf{x}_t, t)$ is Tweedie's denoised prediction and $\mathbf{H}_p^{\dag}$ is a pseudo-inverse of $\mathbf{H}_p$. In this paper, we focus on evaluating how good these approximations are when the posterior is not peaked or even multimodal.\vspace{-0.5cm}}\label{tab:grad}
\end{table}
\section{Posterior evaluation}\label{sec:method}
 \noindent In this section, we introduce two properties that will allow us to compare the approximate posterior, \ie samples from a PnP diffusion method, to the true unknown posterior.
 \\
\textbf{Proposition - }
 Let $p_{\text{Y}_p}(\mathbf{y}_p)$ be the density of $p$-projections sinograms and $p_{\text{X} \vert \text{Y}_p}(\mathbf{x} \vert \mathbf{y}_p)$ the true posterior. Then,
\begin{align}
    \label{eq:Gen}\mathbb{E}_{\mathbf{y}_p \sim p_{\text{Y}_p}}&[p_{\text{X} \vert \text{Y}_p}(\mathbf{x} \vert \mathbf{y}_p)] = p_{\text{X}}(\mathbf{x}),\\
    \frac{1}{m_p \sigma_y^2}\mathbb{E}_{\mathbf{y}_p \sim p_{\text{Y}_p}}&\mathbb{E}_{{\mathbf{x}} \sim p_{\text{X} \vert \text{Y}_p}}\left[\|\mathbf{y}_p - \mathbf{H}_p\mathbf{x}\|_2^2\right] = 1,\label{eq:MCp}
\end{align}

where Eq.~\eqref{eq:MCp} is derived from the result $\mathbb{E}_{\mathbf{y}_p \sim p_{\text{Y}_p \vert \text{X}}}\left[\|\mathbf{y}_p - \mathbf{H}_p\mathbf{x}\|_2^2\right] = m_p \sigma_y^2$. Let us highlight that Eq.~\eqref{eq:Gen} and Eq.~\eqref{eq:MCp} are not dependent on the number of projections $p$. The left hand side of both equations requires evaluating an expectation over $p_{\text{Y}_p}$. This can be done empirically by successively sampling from the prior $p_{\text{X}}$ and the likelihood $p_{\text{Y}_p \vert \text{X}}$.

In practice, the true posterior $p_{\text{X} \vert \text{Y}_p}$ is unknown, but a good approximate posterior $\overset{\sim}{p}_{\text{X} \vert \text{Y}_p}$ is expected to exhibit the same properties. In this context, replacing $p_{\text{X} \vert \text{Y}_p}$ with $\overset{\sim}{p}_{\text{X} \vert \text{Y}_p}$ in Eq.~\eqref{eq:Gen} states that a set of approximate posterior samples (where the conditioning sinograms are sampled from $p_{\text{Y}_p}$, see above) $\{\hat{\mathbf{x}}_{p}^{(i)}\}_{i=1}^{N}$ should not be distinguishable from a set of prior samples $\{\mathbf{x}^{(i)}\}_{i=1}^{N}$.
Therefore, any classical metric computing a distance between two sets of samples, such as FID, KID or CMMD~\cite{jayasumana2024rethinking}, should be close to zero and independent of the number of projections $p$. We refer to this distance as Posterior-Prior Similarity (PPS), and $\text{PPS}_{\text{FID}}$ and $\text{PPS}_{\text{CMMD}}$ as the implementation of PPS using FID and CMMD, respectively.

However, this criterion alone is not sufficient to evaluate the gap between $p_{\text{X} \vert \text{Y}_p}$ and $\overset{\sim}{p}_{\text{X} \vert \text{Y}_p}$, as any method essentially ignoring the conditioning sinogram $\mathbf{y}_p$ and sampling from the prior would satisfy it.
To circumvent this limitation, we also evaluate the second criterion Eq.~\eqref{eq:MCp}, that can be computed as follows: $\sum_{i=1}^{N}\| \mathbf{y}_{p,i} - \mathbf{H}_p\hat{\mathbf{x}}_{p,i}\|_2^2 / ({N} m_p \sigma_y^2)$ where $m_p$ is the dimension of sinograms and the expectation over the posterior is reduced to a single posterior sample. In the remainder of the paper, this quantity is referred as the Normalized average Measurement Consistency (NMC). Let us highlight that the NMC should be close to 1. A value close to 0 indicates that the method only ``transforms" the conditioning sinogram into an image without denoising it. A value significantly larger than 1 reveals that the method essentially ignores the conditioning sinogram.

In practice, the gap between the approximate posterior and the true posterior has two origins (besides the SDE discretization):
the approximation $\mathbf{s}_{\boldsymbol{\theta}}(\mathbf{x}_t, t) \approx \nabla_{\mathbf{x}_t}\log p_t(\mathbf{x}_t)$, and the approximation of $\nabla_{\mathbf{x}_t}\log p_t( \mathbf{y}_p \vert \mathbf{x}_t)$ (see Tab.~\ref{tab:grad}). As this paper focuses on evaluating the latter, in the experiments we get rid of the first approximation as follows: each time a prior sample is required, we obtain it by sampling (unconditionally) using $\mathbf{s}_{\boldsymbol{\theta}}$. Consequently, in the experiments, $\mathbf{s}_{\boldsymbol{\theta}}$ is no longer an approximation but the exact time-dependent prior score function, and only the approximation of $\nabla_{\mathbf{x}_t}\log p_t( \mathbf{y}_p \vert \mathbf{x}_t)$ is evaluated. Another advantage of this procedure is that an arbitrary large number of prior images can be sampled, which is very important for the FID, that needs to be computed on large sets of images to be unbiased~\cite{chong2020effectively, jayasumana2024rethinking}.

\begin{table*}[h!]
    \scriptsize
    \begin{tabularx}{1.0\textwidth}{D{0}D{0.4} D{0.5}D{0}D{0.4}D{0.5} D{0.3} D{0.5}D{0}D{0.4}D{0.5} D{0.3} D{0.5}D{0}D{0.4}D{0.5}}
    \specialrule{.1em}{1em}{0em} \\
    \multicolumn{2}{c}{} & \multicolumn{4}{c}{MCG \cite{chung2022improving}} && \multicolumn{4}{c}{DPS \cite{chung2023diffusion}} && \multicolumn{4}{c}{$\Pi$G \cite{song2023pseudoinverse}}\\
    \cline{3-6} \cline{8-11} \cline{13-16}

    \\

     & &  \multicolumn{1}{c}{Eq. \eqref{eq:MCp}} && \multicolumn{2}{c}{Eq. \eqref{eq:Gen}} &&       \multicolumn{1}{c}{Eq. \eqref{eq:MCp}} && \multicolumn{2}{c}{Eq. \eqref{eq:Gen}} &&       \multicolumn{1}{c}{Eq. \eqref{eq:MCp}} && \multicolumn{2}{c}{Eq. \eqref{eq:Gen}}\\
     \cline{3-3}\cline{5-6}\cline{8-8}\cline{10-11}\cline{13-13}\cline{15-16}\\
     
     & $p$ &  $\text{PLS}_{\text{NMC}}$ && $\text{PPS}_{\text{FID}}$ & $\text{PPS}_{\text{CMMD}}$ & &  $\text{PLS}_{\text{NMC}}$ && $\text{PPS}_{\text{FID}}$ & $\text{PPS}_{\text{CMMD}}$ & &  $\text{PLS}_{\text{NMC}}$ && $\text{PPS}_{\text{FID}}$ & $\text{PPS}_{\text{CMMD}}$\\
     & & $\xrightarrow{} 1$ && $\downarrow$ & $\downarrow$ & &  $\xrightarrow{} 1$ && $\downarrow$ & $\downarrow$ & &  $\xrightarrow{} 1$ && $\downarrow$ & $\downarrow$\\
     \midrule
     \multirow{9}{*}{\rotatebox[origin=c]{90}{LDCT}} 
     & $180$  &  $\mathbf{0.89}$ && $\mathbf{0.57}$ & $\mathbf{0.002}$ && $1.14$ && $0.58$ & $0.003$ && $-$ && $-$& $-$\\
      &$90$ &  $0.82$ && $1.06$ & $0.088$ && $\mathbf{1.12}$ && $\mathbf{0.57}$ & $\mathbf{0.003}$&& $-$ && $-$& $-$\\
      &$30$ & $0.65$ && $2.27$ & $0.386$ && $\mathbf{1.05}$ && $\mathbf{0.54}$ & $\mathbf{0.001}$ && $-$ && $-$& $-$\\
      &$18$ &  $0.45$ && $2.61$ & $0.449$ && $\mathbf{0.98}$ && $\mathbf{0.55}$ & $0.004$ && $1.92$ && $\mathbf{0.55}$& $\mathbf{0.0012}$\\
      &$12$ & $0.32$ && $3.90$ & $0.611$ && $\mathbf{0.90}$ && $0.61$ & $0.012$ & & $2.01$ && $\mathbf{0.54}$& $\mathbf{0.0003}$\\
       &$6$ & $0.20$ && $5.94$ & $0.733$ && $\mathbf{0.72}$ && $0.82$ & $0.047$ && $2.19$ && $\mathbf{0.54}$& $\mathbf{0.0003}$\\
       &$3$ & $0.15$ && $7.66$ & $0.829$ && $\mathbf{0.55}$ && $1.40$ & $0.125$ && $3.02$ && $\mathbf{0.55}$& $\mathbf{0.0002}$\\
       &$1$ & $0.20$ && $8.54$ & $0.754$ && $\mathbf{0.24}$ && $6.34$ & $0.563$ && $14.89$ && $\mathbf{1.00}$ & $\mathbf{0.0070}$\\ 
     \midrule
     \multirow{9}{*}{\rotatebox[origin=c]{90}{LIDC}} 
     & $180$  & $\mathbf{0.87}$ && $1.34$ & $0.022$ && $1.48$ && $\mathbf{0.87}$ & $\mathbf{0.012}$ && $-$ && $-$ & $-$\\
      &$90$ & $\mathbf{0.80}$ && $1.00$ & $0.009$ && $1.45$ && $\mathbf{0.86}$ & $\mathbf{0.012}$ && $-$ && $-$ & $-$\\
      &$30$ & $0.63$ && $2.35$ & $0.166$ && $\mathbf{1.35}$ && $\mathbf{0.84}$ & $\mathbf{0.010}$ && $-$ && $-$ & $-$\\
      &$18$ & $0.46$ && $3.25$ & $0.278$ && $\mathbf{1.20}$ && $0.89$ & $0.010$ && $3.01$ && $\mathbf{0.54}$ & $\mathbf{0.0007}$\\
      &$12$ & $0.34$ && $4.44$ & $0.360$ && $\mathbf{1.04}$ && $1.01$ & $0.012$ && $3.79$ && $\mathbf{0.53}$ & $\mathbf{0.0005}$\\
       &$6$ & $0.22$ && $5.67$ & $0.411$ && $\mathbf{0.74}$ && $1.20$ & $0.019$ && $4.71$ && $\mathbf{0.53}$ & $\mathbf{0.0005}$\\
       &$3$ & $0.15$ && $5.97$ & $0.421$ && $\mathbf{0.51}$ && $1.39$ & $0.031$ && $7.12$ && $\mathbf{0.56}$ & $\mathbf{0.0008}$\\
       &$1$ & $0.21$ && $4.90$ & $0.337$ && $\mathbf{0.27}$ && $2.80$ & $0.164$ && $14.58$ && $\mathbf{0.67}$ & $\mathbf{0.0019}$\\
       \midrule
        \multirow{9}{*}{\rotatebox[origin=c]{90}{JET}} 
        & $180$  & $\mathbf{0.98}$ && $4.87$ & $0.151$ && $1.70$ && $\mathbf{1.03}$ & $\mathbf{0.040}$ && $-$ && $-$ & $-$\\
      &$90$ & $\mathbf{0.96}$ && $4.86$ & $0.086$ && $1.68$ && $\mathbf{1.01}$ & $\mathbf{0.040}$ && $-$ && $-$ & $-$\\
      &$30$ & $\mathbf{0.90}$ && $2.29$ & $0.090$ && $1.63$ && $\mathbf{0.96}$ & $\mathbf{0.042}$ && $-$ && $-$ & $-$\\
      &$18$ & $\mathbf{0.81}$ && $5.06$ & $0.542$ && $1.51$ && $1.02$ & $0.044$ && $2.11$ && $\mathbf{0.62}$ & $\mathbf{0.008}$\\
      &$12$ & $0.63$ && $10.23$ & $1.096$ && $\mathbf{1.36}$ && $1.28$ & $0.060$ && $5.25$ && $\mathbf{0.64}$ & $\mathbf{0.009}$\\
       &$6$ & $0.30$ && $14.20$ & $1.733$ && $\mathbf{1.07}$ && $2.23$ & $0.141$ && $6.57$ && $\mathbf{0.63}$ & $\mathbf{0.009}$\\
       &$3$ & $0.17$ && $14.21$ & $2.298$ && $\mathbf{0.83}$ && $3.80$ & $0.235$ && $7.99$ && $\mathbf{0.58}$ & $\mathbf{0.007}$\\
       &$1$ & $0.21$ && $13.96$ & $1.941$ && $\mathbf{0.48}$ && $12.89$ & $1.061$ && $9.72$ && $\mathbf{0.56}$ & $\mathbf{0.005}$\\
       \bottomrule
    \end{tabularx}
	\caption{Posterior evaluation of three state-of-the-art PnP diffusion methods \cite{chung2022improving, chung2023diffusion, song2023pseudoinverse} in SVCT. See Sec.~\ref{sec:quantitative_eval} for details. Corrected $\text{PLS}_{\text{NMC}}$ results following the identification of a computational bug in the official publication. The $\text{PLS}_{\text{NMC}}$ corrected values presented here replace those in the original table, which contained errors due to the bug. These corrected results do not affect the overall conclusions of the study.}
	\label{tab:tableEval}
\end{table*}

\subsection{Training Datasets}
\noindent We consider three datasets:\\
1 - The \jetOnera\footnote{The dataset will be made available.} dataset~\cite{leon22dhi} contains $5896$ compressible flow images of size $260\times 260$ (see Fig.~\ref{fig:JetSamples}). 
These images are cross-sections of hot air (compressible flow density) expelled by a nozzle and impacting a wall at a given distance~\cite{grenson2017jet}.\\ 
2 - The Low Dose CT grand challenge $2016$ (LDCT)~\cite{mccollough2016tu} contains $5936$ CT images of size $512\times 512$ from $10$ patients. We use the 1mm thick scans. The scans are acquired with the same scanner and hyperparameters.\\
3 - The Lung Image Database Consortium (LIDC) dataset~\cite{armato2011lung, clark2013cancer} contains $239472$ CT images of size $512\times 512$ from $1018$ patients. Unlike LDCT scans, LIDC scans are acquired with different scanners and different hyperparameters (\textit{e.g.} thickness, tube peak potential energies, tube current range, \textit{etc.}).\\
In the following experiments, all the images
are resized to $128\times 128$ and normalized between $[0, 1]$.
Sinograms are obtained using the \textit{torch-Radon} library~\cite{ronchetti2020torch}. The noise level added to sinograms is determined on each dataset by taking 1\% of the dataset's 180-projections sinograms average dynamic:
\begin{equation}
    \sigma_y = \frac{1}{100 \times {N}}\sum_{i\in\mathcal{D}}\left(\text{max}(\mathbf{H}_{180}\mathbf{x}_i) - \text{min}(\mathbf{H}_{180}\mathbf{x}_i)\right).
\end{equation}
\subsection{Prior score network}
\noindent For each dataset we train a time-dependent prior score network $\mathbf{s}_{\boldsymbol{\theta}}$. We use a \texttt{ncsnpp} architecture, \ie a Unet augmented with attention, in VE-SDE continuous settings with $\sigma_{min} = 0.01$ and $\sigma_{max} = 1348$ similarly to the Pytorch implementation of~\cite{song2021score}. The same model is shared between the different PnP diffusion methods to ensure comparable results. On the \jetOnera dataset, the images are augmented with random horizontal and vertical flips and random rotation. We trained $\mathbf{s}_{\boldsymbol{\theta}}$ on a single NVIDIA TITAN~V $12$Go, using Adam optimizer with a learning rate of $10^{-4}$.

\subsection{Technical details on PnP diffusion methods}\label{sec:technical_details_pnp}
\noindent We consider three SOTA PnP diffusion methods: MCG~\cite{chung2022improving}, DPS~\cite{chung2023diffusion} and $\Pi$G~\cite{song2023pseudoinverse}. They essentially differ in the respective approximations of $\nabla_{\mathbf{x}_t}\log p_t( \mathbf{y}_p \vert \mathbf{x}_t)$ (see Tab.~\ref{tab:grad}).
MCG and DPS have handcrafted weights originally defined as $\alpha_{\text{MCG}}(\mathbf{x}_t, \mathbf{y}_p)=0.1/\|\mathbf{H}_p^{\dag}(\mathbf{y}_p - \mathbf{H}_p\hat{\mathbf{x}}_0(\mathbf{x}_t))\|_2$ and $\alpha_{\text{DPS}}(\mathbf{x}_t, \mathbf{y}_p) = 1/\|\mathbf{y}_p - \mathbf{H}_p\hat{\mathbf{x}}_0(\mathbf{x}_t)\|_2$.  We found these values to be non-optimal and fine-tuned them on each dataset for each value of $p$. $\Pi$G proposes time-dependent weights: $r_t = \sqrt{\sigma_t^2 / \sigma_t^2 + 1}$ with $\sigma_t$ the diffusion noise, we did not modify this scheme. To fairly compare the three methods, in all the experiments, we use the same accelerated ancestral sampler\cite{song2021denoising, song2023pseudoinverse} with 100 noise scales with $\sigma_{min} = 0.01$ and $\sigma_{max} = 1348$.

To improve the numerical stability of MCG and DPS, we implemented the gradient of the time-dependent likelihood term with its closed-form expression (see Tab. \ref{tab:grad}).

\subsection{Quantitative evaluation}\label{sec:quantitative_eval}
\noindent We aim at evaluating the approximate posterior of each PnP diffusion method~\cite{chung2022improving, chung2023diffusion, song2023pseudoinverse} as a function of the number of projections $p$ (in the experiments $p=1,3,6,12,18,30,90,180$). To do so, for each dataset, we proceed as follows. We unconditionally sample a set $\mathcal{X}$ of $50$k images using the time-dependent prior score network $\mathbf{s}_{\boldsymbol{\theta}}$, as explained in Sec.~\ref{sec:method}. From $\mathcal{X}$, for each value of $p$, we compute a set of $50$k $p$-projections sinograms, $\mathcal{Y}_p$, using the SVCT measurement model Eq.~\eqref{eq:observation_model}. Then for each method $m\in$~\cite{chung2022improving, chung2023diffusion, song2023pseudoinverse}, we compute a set of $50$k posterior samples, $\overset{\sim}{\mathcal{X}}_{p,m}$, using sinograms of $\mathcal{Y}_p$ as conditions. To evaluate Eq. \eqref{eq:Gen} we compute the FID~\cite{heusel2017gans} and the CMMD~\cite{jayasumana2024rethinking} between $\mathcal{X}$ and $\overset{\sim}{\mathcal{X}}_{p,m}$. To evaluate Eq. \eqref{eq:MCp}, we compute the NMC (see Sec.~\ref{sec:method}) between $\mathcal{Y}_p$ and 
$\overset{\sim}{\mathcal{X}}_{p,m}$.
The evaluation results are summarized in Tab.~\ref{tab:tableEval}. Let us highlight that to compute $\nabla_{\mathbf{x}_t}\log p_t( \mathbf{y}_p \vert \mathbf{x}_t)$,
$\Pi$G needs to solve a linear system whose size increases with $p$ (see Tab.~\ref{tab:grad}), thus we could only compute the metrics for $p = \{1, 3, 6, 12, 18\}$. In theory MCG needs to perform a similar operation (pseudo-inverse) but, as suggested in~\cite{chung2022improving}, we use a Filtered Back-Projection~\cite{kak2001principles} (FBP), that significantly speeds-up the sampling.

%
\textbf{Results of MCG - } At $p=180$, the NMC is slightly below $1$, indicating a mild overfitting regime. As $p$ decreases, MCG shows increasing overfitting, with the NMC approaching $0$. Both the FID and CMMD metrics rise significantly as $p$ decreases, which aligns with the NMC results. Indeed, greater overfitting leads to more correlated noise in the posterior samples. When this correlated noise is projected back into sinogram space, it manifests as the overfitted noise and hence differs from prior samples. However, the observed dependency on $p$ is unexpected, indicating that MCG’s approximate posterior begins to diverge from the true posterior as $p$ decreases.\\
%
%
\textbf{Results of DPS - } At $p=180$, the NMC is slightly over $1$, indicating a mild underfitting regime. As $p$ decreases, DPS transitions into overfitting regime.Both the FID and CMMD metrics increase significantly in the overfitting scenario. This is again due to the overfitting correlated noise in posterior samples. The NMC and FID, CMMD seems to be ideal around $p=18$ but we still observe a dependency in $p$.\\
\textbf{Results of $\boldsymbol{\Pi}$G - } The FID and the CMMD are very low, but the NMC is quite high and far from 1. This shows $\Pi$G does not take correctly into account the conditioning sinogram but produces samples almost indistinguishable  from prior samples. This behaviour may be explained by $\Pi$G's dynamic weighing scheme $r_t$ (see Sec.~\ref{sec:technical_details_pnp}) which does not correctly balance the two terms of Eq.~\eqref{eq:Bayes}.

This quantitative evaluation shows that DPS's approximate posterior is much closer to the true posterior than the approximate posteriors of MCG and $\Pi$G. This finding is unexpected as DPS employs a simpler and faster approximation of $\nabla_{\mathbf{x}_t}\log p_t( \mathbf{y}_p \vert \mathbf{x}_t)$ than MCG and $\Pi$G. 

\section{Conclusion}
\noindent In this work, we introduced two posterior evaluation criteria to assess the approximate posterior of a PnP diffusion method. To the best of our knowledge, this is the first attempt to evaluate SVCT posteriors with generative metrics, and in a non-peaked regime. This lead to a thorough quantitative evaluation of three state-of-the-art methods on three different datasets. Our findings surprisingly show that the approximate posterior tends to deviate from the true posterior when the number of projections decreases. Moreover, the best posterior approximation is obtained by the simplest and fastest method DPS. These findings would not have been possible using traditional image-to-image metrics like PSNR and SSIM.

\bibliographystyle{IEEEbib}
\bibliography{refs}

\end{document}